\documentclass[aps, pre, preprint, longbibliography,floatfix, nofootinbib, superscriptaddress]{revtex4-1}
\usepackage{amsmath,amsfonts}
\usepackage{graphicx}
\usepackage{color}
\usepackage{subfigure}
\usepackage{tikz}
\usepackage{tikz-3dplot}
\usetikzlibrary{calc,quotes,angles}
\usepackage[colorlinks=true, citecolor=red, linkcolor=blue,urlcolor=blue ]{hyperref}

\usepackage{amssymb}
\usepackage{hyperref}
\usepackage{subfigure}
\usepackage{url}

\usepackage{natbib}

\begin{document}

\title{Fractal dimension and topological invariants as methods to quantify complexity in Yayoi Kusama's paintings}

 \date{\today}
 \author{Elsa \surname{de la Calleja}}
\affiliation{Instituto de Investigaciones en Materiales, Facultad de Ingenier\'ia, Universidad Nacional Autónoma de M\'exico, Av. Universidad 3000, Ciudad de M\'exico, 04510, M\'exico}
\author{Roberto \surname{Zenit}}
\email{zenit@brown.edu}
 \affiliation{Instituto de Investigaciones en Materiales, Facultad de Ingenier\'ia, Universidad Nacional Autónoma de M\'exico, Av. Universidad 3000, Ciudad de M\'exico, 04510, M\'exico}
 \affiliation{School of Engineering, Brown University, 184 Hope St, Providence, RI 02912, USA}

\begin{abstract}
Intricate patterns in abstract art many times can be wrongly characterized as being complex. Complexity can be an indicator of the internal dynamic of the whole system, regardless of the type of system in question, including art creation. In this investigation, we use two different techniques to objectively quantify complexity in abstract images: the fractal dimension and the value of the Betti numbers.  We first validate our technique by considering synthetic images with a random distribution of dots, to then apply it to a series of `Net obsession' paintings by Yayoi Kusama. Surprisingly, we found that although the fractal dimension of her works in this series is comparable to those by Jackson Pollock in his dripping period, which could indicate a high level of complexity, the value of the Betti numbers do show disconnectedness and not high complexity. This is intuitively in agreement with the visual assessment of such works.
\end{abstract}


\maketitle


\section{Introduction}

The symmetry, complexity or periodicity of irregular spatial-temporal patterns could be characterized as beautiful. For many years, numerous investigations have tried to characterize the dynamical processes that lead to the creation of such patterns.  Structural properties in many complex systems are the final manifestations of the internal dynamics. Through measures such as connectivity, porosity, texture, roughness, symmetry or temporal correlations to name a few, the complexity of systems can be characterized \cite{Ball:2016dg,Barnsley:1993dg,Benoit:1982qr,Etzenhouser:1998jd,Daccord:1986jd,Goodchild:1987jd}.
Nature produces many examples of  complex patterns \cite{Caballero:2012jd,Barnsley:1993dg,Andrienko:2018jd}. Some  investigations have tried to described the dynamics behind natural  complexities, many of which are visual: for example Turing patterns \cite{Caballero:2012jd}, symmetries, textures and patterns of marine invertebrates \cite{Ball:2016dg}, etc.

Art, of course, is full of complexity but its objective quantification is not trivial. Some studies have quantified particular characteristics of abstract paintings. For instance, Naumis \emph{et al.} \cite{Naumis:2008jd} evaluated the \emph{turbulent} luminance of van Gogh paintings, Mekler \emph{et al.} \cite{Mekler:2009jd} measured size-ordered distributions in Vasili Kandinsky. The physical process leading to the emergence of complex  patterns in the David A. Siqueiros accidental painting technique \cite{Zenit:2019dg} and in Jackson Pollock's drip paintings \cite{Zenit:2015dg} have been studied by our group. The fractal dimension of artworks has been widely used to assess the degree of complexity \cite{Mureika:2013dg,Taylor:1999dg,Elsama:2016dg,Redies:2007dg,Zenit:2019dg,Zenit:2015dg}. This attempt has created a renewed interest in the subject and opens a new area of study \cite{Abbot:2006dg,Coddinton:2008dg}, even if their validity and extend of applicability have been questioned \cite{Jones:2006,Jones:2009}. Despite these efforts, a general consensus on the appropriate  mathematical tools to quantify the complexity of abstract art has not been reached. 

In this report, the complexity of an image is evaluated using those two different mathematical approaches: fractal dimension and Betti numbers. Both measures are used to characterize structural properties as geometry, symmetry, non-regularity, dimensionality and connectivity of a set. Felix Hausdorff developed the mathematical description for Mandelbrot's sets better known as  fractals \cite{Benoit:1982qr}. It is well accepted that the Hausdorff-Besicovitch fractal dimension is the usual parameter to characterize irregular spatial-temporal patterns as a result of temporal or spatial evolution of the elements of the system \cite{Mecke:1996jd,Ball:2016dg,Forsythe:2011}. On the other hand, the modern formulation of the theory of algebraic topology \cite{Lane:1985jd} is also a powerful tool to analyze complexity in many  systems by using the value of topological invariants, the so-called Betti numbers \cite{Pranav:2017jd,Edgar:1990qr,DelaCalleja:2017,Elsama:2016dg}. Despite being widely used in many other areas to quantify structural properties \cite{Kosterlitz:1973,Gottsche:1990dg,Spanslatt:2005dg,Kondic:2012dg,Xia:2014dg}, the use of topological invariants as a measure of complexity has not been used much for the case of art. Recently, the Betti numbers of abstract expressionist artworks were computed and used as a measure of complexity \cite{DelaCalleja:2017}, with promising results to be applied in other cases. 

To validate our proposed methodology to quantify complexity we first apply it to sets of black and white synthetic abstract images formed by randomly adding adding dots (or lines) in a two-dimensional empty area. The advantage of using black-white two-dimensional images allows applying the technique without any previous image processing. Then, the same methodology was used to measure the complexity in Yayoi Kusama's artworks. Her work is characterized by the use of regular dots patterns. Some critics have cataloged such periodic distribution of dots and  nets as `simple'. The analysis of Kussama's works is of particular importance because we found that their fractal dimension hold similar values as other abstract paintings, such as Pollock's; however, the value of the Betti numbers indicate a lack of connectivity and not necessarily complexity. Hence, we show that the fractality of an image cannot be used alone to quantify complexity. Furthermore, our results indicate that complexity is not related to the visual appreciation of the image. 

\section{Mathematical parameters to quantify complexity}
To investigate the complexity in abstract images, two methods were considered to assess the structural properties of images: the fractal dimension and the value of topological invariants. Both procedures have been used in many cases from spatial distribution of galaxies to music \cite{Pranav:2017jd,Barnsley:1993dg,Bigerelle:2000jd}. In this study these mathematical parameters were calculated for a set of synthetic abstract images composed  with dots or lines. As a result, two different measures were obtain for a given image. This allows us to compare and contrast the structural properties of the images. Evidently, in the art world the painting composition contains multiple components, such as dense lines, layers of paint overlay, cluster nodes, voids, or shapes and shadows. And many of those objects live in a three dimensional space. 

\subsection{The Betti numbers, $\beta_i$}
To sustain the characterization of the connectivity of abstract images using topological invariants, a brief introduction is presented here. Algebraic topology is the branch of mathematics that studies topological spaces \cite{Milnor:1963dg,Munkres:1984dg,Edelsbrunner:2010dg}. The formalism allows the formulation of statements about topological spaces into the language of group theory, offering substantial flexibility and a deeper understanding of spatial structure and connectivities \cite{Pranav:2017jd}.  A topological space is characterized by its invariants, which are numbers that remain unchanged under a homeomorphic transformation \cite{Edgar:1990qr,Spanier:1966qr}. All homology groups are vector spaces, their dimension is the number of independent $i-$dimensional cycles in a topological space. This is the formal definition for the Betti numbers $\beta_i$, where $i=0,1,2,...,d$ \cite{Betti:1987jd,Edelsbrunner:2010dg}. The Betti numbers provide the general description of structural topology in systems or sets where the mass distribution play a relevant role.

The description of the boundaries of holes \cite{Munkres:1984dg} allow the characterization of the connectivity of set in a space. For a two dimensional space, only the two Betti numbers can be defined: $\beta_{0}$ measures the number of simply-connected objects, path-connected or isolated components of \emph{X}; and $\beta_{1}$ counts the number of loops enclosing independent tunnels, it means the number of holes within the simply-connected spaces or the $\mu$-dimensional holes in \emph{X}. By taking into consideration this reduced definition, we used the values of these two numbers to assess to the topological properties on abstract images and quantify their complexity.

\subsection{The fractal dimension}
A fractal object can be considered as an irregular set taking by reference the irregularity in classical geometry. No matter how much the set is magnified, smaller and smaller irregularities become visible \cite{Benoit:1982qr,Edgar:1990qr}. Roughly speaking, there are similar features on fractal objects: \emph{(i)} self-similarity; \emph{(ii)} recursive procedure for its construction; \emph{(iii} its size is not quantified by the usual measures such as length; \emph{(iv)} the object has a fine structure, which contains details at arbitrarily small scales; and some others~\cite{Falconer:2014qr}. Taking into account the above characteristics the fractal dimension is not a topological invariant property, since fractal objects are not homeomorphic \cite{Benoit:1982qr,Edgar:1990qr,Harewicz:1941qr,TaylorP:1985jd,Taylor2:1986jd,Falconer:1987jd,Hilton:1988jd}. Mandelbrot~\cite{Benoit:1982qr} defined a fractal as a set in which the Hausdorff-Besicovitch dimension exceed the topological dimension.

Briefly, the formalism used consist of given a non-empty space of n-dimensional Euclidean space, the box counting fractal dimension can be calculated by the general equation $N(S)=(\epsilon/\epsilon_0)^{-f(\alpha)}$ for small $\epsilon$ where $N(S)$ is the number of boxes used to cover the set $S$, $\epsilon$ is the size of the box and $\epsilon_0$ as the minimum size of box. The measure of fractal dimension is compute with a digital binarized image which is divided into pieces of size $\epsilon$. Halsey et al. \cite{Halsey:1986dg} defined $P_{i,Q}^{Q}\sim\epsilon_{i}/\epsilon_0^{\alpha Q}$ as the probability to visit a site in different scaling indices $Q$, specifically, the number of times that $\alpha$ in $P_{i,q}$ takes a value between $\alpha'$ and $d\alpha'$ defined as $d\alpha'\rho(\alpha')\epsilon^{-f(\alpha')}$ where $f(\alpha')$ is a continuous function. The fractal dimension for all the scaling indices can be evaluated 
\begin{equation}
D_{Q}=\frac{1}{1-Q}\lim_{\epsilon\rightarrow 0}\frac{ln \sum_{i=1}^{N(\epsilon)}[P_{(i,Q)}]^{Q}}{ln (\epsilon/\epsilon_0)}
\end{equation}
As \emph{Q} represents different scaling indices, we can define
\begin{equation}
\sum_{i=1}^{N(\epsilon)}[P_{(i,Q)}]^{Q}=\int d\alpha'\rho(\alpha')\epsilon^{-f(\alpha')+Q\alpha'}
\end{equation}
where the  Lipschitz-H\"{o}lder exponent $\alpha_{i}$, characterizes the singularity strength in the \emph{ith} box and quantifies the distribution of complexity in an spatial location. There are similar definitions of fractal dimension, and several techniques to measure it. To mention a few: the box-counting method~\cite{Barnsley:1993dg}, Hausdorff measure \cite{Falconer:2014qr}, Packing dimension \cite{Edgar:1990qr,Xiao:1997jd}, fractal spectrum \cite{Chhabra:1989jd} on some others. Nevertheless, the methods are based on equivalent definitions of fractal dimension.  In this report the fractal spectrum with the box-counting method is used to quantify the fractality on the sets \cite{Chhabra:1989jd}.

\section{Methods and Methodology}
\begin{figure}[!ht]
\centering
\includegraphics[width=0.75\linewidth]{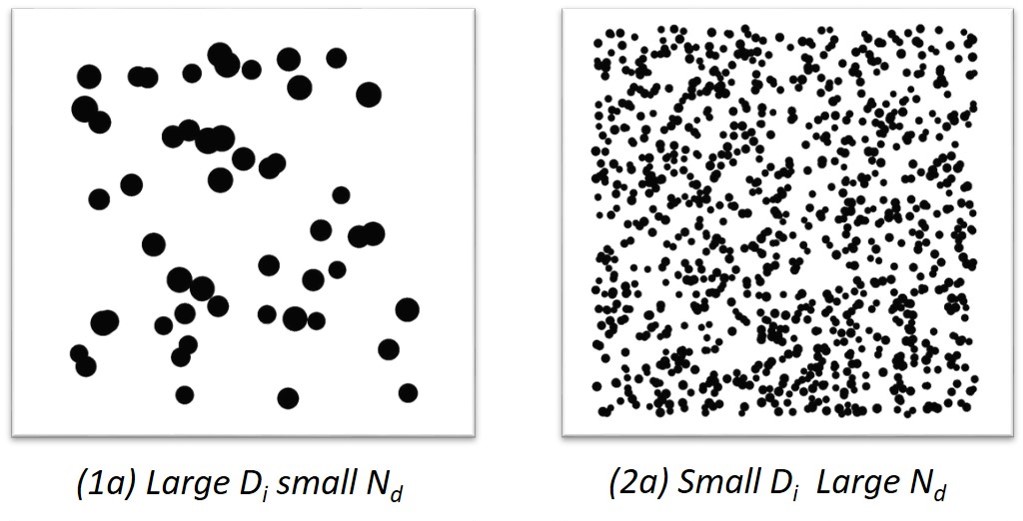}
\caption{Typical synthetic abstract images generated with random distribution of dots.}
\label{Image:example-dots}
\end{figure}

In this  work we evaluate the complexity of synthetic abstract images considering the values of the  fractal dimension and the Betti numbers. The procedure to calculate complexity was the following: series of synthetic abstract images were produced using a script written with a Matlab$\copyright$. Two kind of images were tested:  dots and lines. In both cases a predetermined number of black objects (dots or lines) were placed randomly within a rectangular blank space. To ensure repeatability, at least three images were produced for each condition. By progressively adding dots or lines on the given area, the black area increases as dots or lines overlap. The area size was kept fixed for all cases. For the case of dots, three series were constructed with dots of three mean diameters: $D_1=0.01$, $D_2=0.02$ and $D_3=0.03$. The dot size distribution has a standard deviation of 10\% of the mean diameter. For the case of lines objects of fixed length, $L=0.1$, were placed randomly in the space with a random orientation. Three series were generated by changing the thickness of the lines, obtained series of $T_1=0.1, T_2=0.3$, and $T_3=0.5$.  The mass distribution of dots or lines on the blank space did not have any distribution preference,  hierarchical aggregation or imposed self-similarity. Typical images generated with dots are shown in Fig.\ref{Image:example-dots}. 

\subsection{Measuring complexity}
The synthetic abstract images were binarized with the use of the free software ImageJ \cite{Schneider:2012dg}. The fractal dimension of each image was calculated with the FracLac plug-in \cite{Chhabra:1989jd,Meneveau:1989jd}. The software calculates the spectrum $f(\alpha)$ of an ordinary uniform set which is a single point on the $f(\alpha)-\alpha$ plane. If the set has fractal characteristics, it shows a line of consecutive points for $Q\geq0$ that starts on the left side of the spectrum climbing up to the maximum value \cite{Halsey:1986dg,Hentschel:1983dg,Schneider:2012dg}. The maximum value for the generalized dimension corresponds to $Q=0$. The maximum value was taken as the fractal dimension. The Betti numbers were measured with the software Chomp~\cite{chomp:2016dg}. The algorithm of this software calculates the local elementary reductions and collapses to compute the homology of the images. The software calculates the first two Betti numbers, $\beta_0$ number, and $\beta_1$ number. The black objects over a blank background can be broadly related to the connectivity among dots or lines and $\beta_1$ number indicates empty spaces corresponding to the number of holes in the image. 

\section{Results for synthetic images}
As mentioned previously, the aim of this report is to analyze the complexity of synthetic abstract images by measuring their fractal dimension and the value of their two Betti numbers.  The case of synthetic images composed by dots is presented and analysed here. The case for lines shows qualitatively similar results; hence, it is presented in Appendix A. By observing the changes in fractal dimension and Betti numbers as the number of dots increases we can evaluate how the complexity evolves. We use `simple' distributions of dots motivated by  our interest to evaluate the complexity of paintings by Japanese avant-garde artist Yayoi Kusama. In 1959, she created \emph{Infinity Nets}, a 10 meters paint  with meticulously inscribed patterns suggestive of a white net, over a slightly darker grayish-white background. Her work has been compared to Jackson Pollock's  \cite{Nakajima:2020dg,Sydney:1958dg}. Visually, it is evident that the two artists' abstract expressionists works are  significantly different. However, as shown below, our measurements of their fractal properties suggest that their paintings have similar structural properties.

\subsection{Fractal dimension of synthetic images}
\begin{figure}
\begin{center}
\subfigure[]{\includegraphics[width=.7\linewidth]{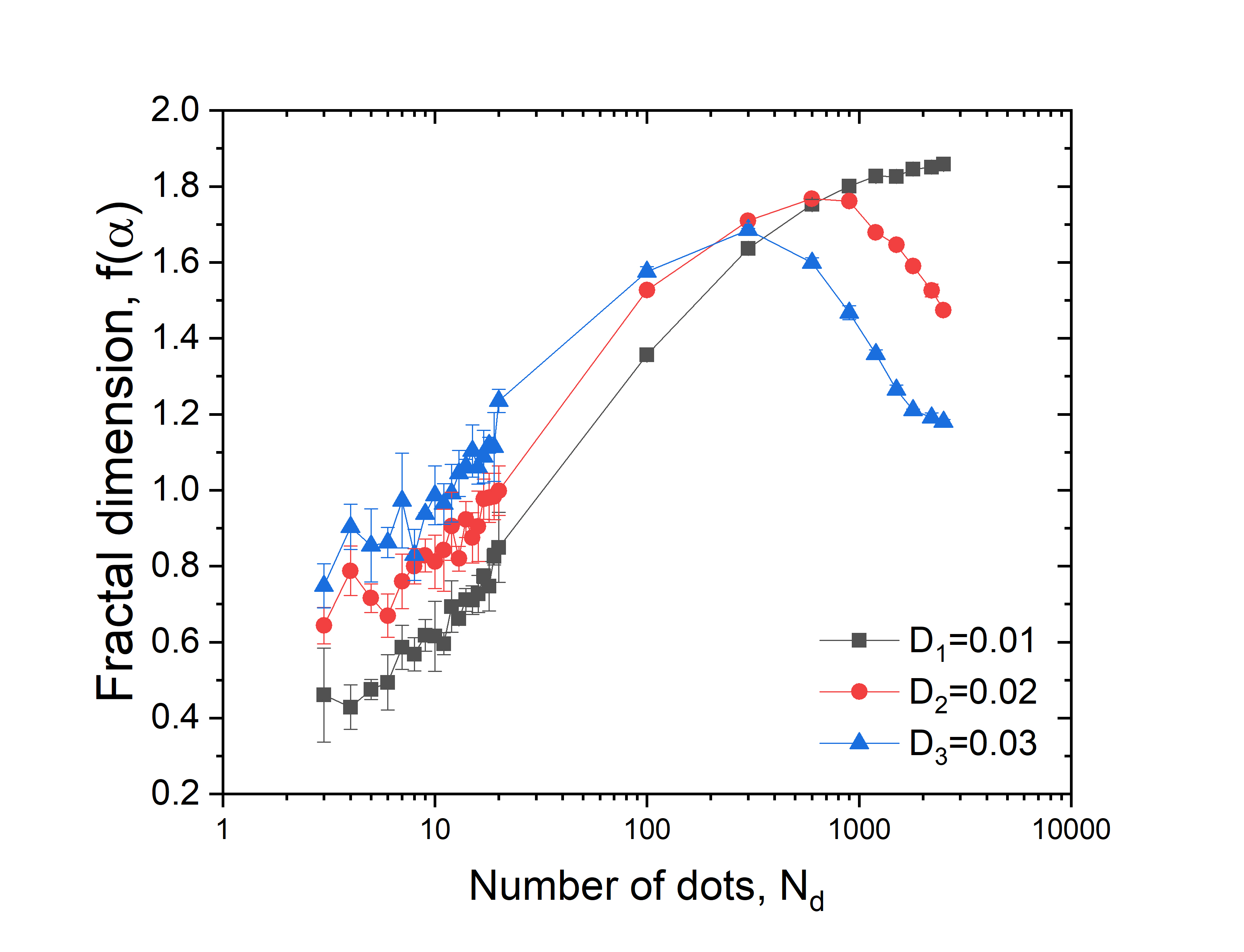}}\\
\subfigure[$f(\alpha)\approx1.5, N_d=100$]{\includegraphics[width=.3\linewidth]{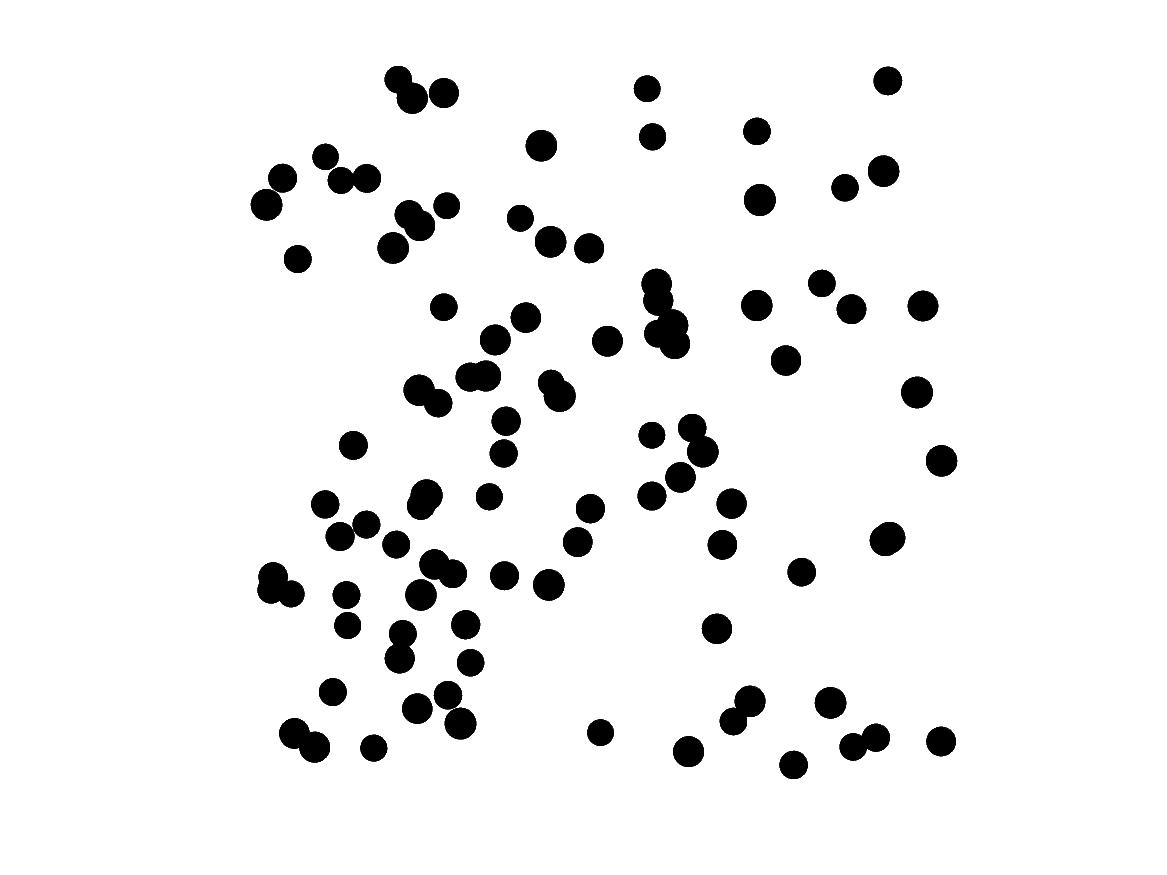}}\,
\subfigure[$f(\alpha)\approx1.7, N_d=500$]{\includegraphics[width=.3\linewidth]{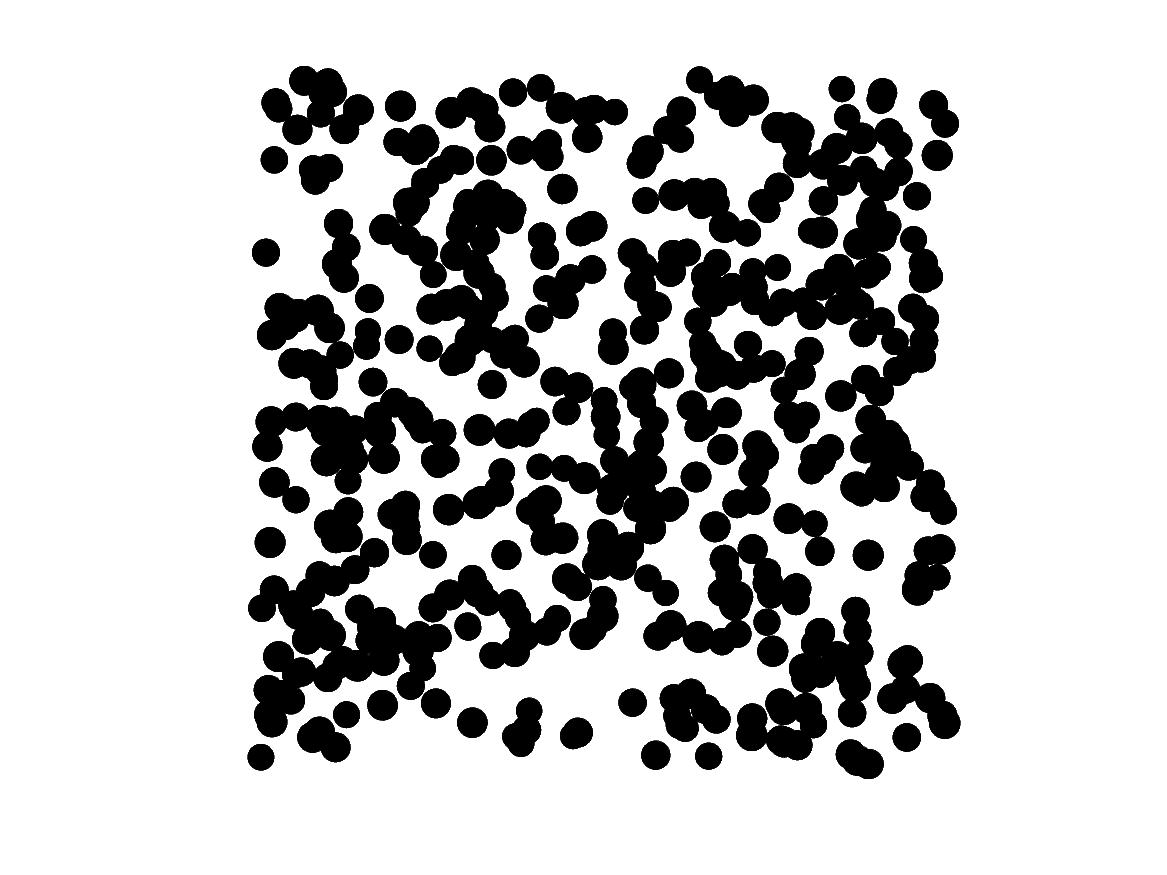}} \,
\subfigure[$f(\alpha)\approx1.5, N_d=2000$]{\includegraphics[width=.3\linewidth]{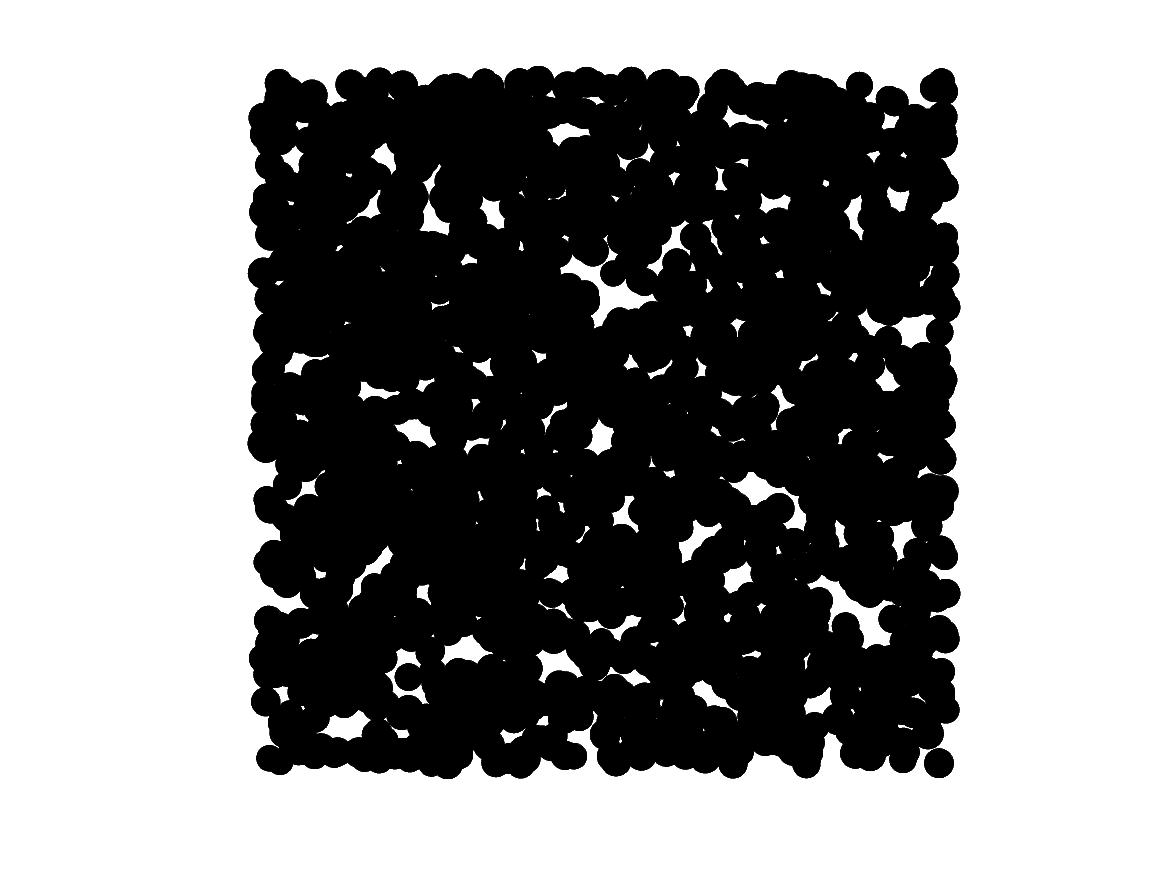}}
\caption{(a) The fractal dimension $f(\alpha)$ as a function of number of dots, $N_d$, for a fixed surface area, for three different drop diameters, $D_1,D_2,D_3$ (black squares, red dots, and blue triangles respectively). (b,c,d) Typical images showing different values $f(\alpha)$, below, at and above the maximum value for $D_2$=0.02.}
\label{Figure:F_D_Dots}
\end{center}
\end{figure}

Figure \ref{Figure:F_D_Dots}(a)  shows the fractal dimension $f(\alpha)$ of the three series of abstract images computed with different sizes $D_1,D_2,D_3$ (black squares, red dots and blue triangles, respectively) as a function of the number of dots, $N_d$. The fractal dimension was measured for images containing from from $2$ to $2500$ dots. Since the area of the image is constant, an increase of the number of dots increases the density. Note also that in these images, the dots are placed randomly and can overlap.  The fractal dimension gradually increases from $0.65$ as the number of dots increases. A maximum fractal dimension of almost $2.0$ is reached at an certain number of dots, which depends on the dot size. These maximum and minimum fractal dimension values are expected \cite{FractGeom2000}. After the fractal dimension reaches a maximum, its value decreases as the number of dots continues to increase.  This is a consequence of the image becoming entirely populated with dots. Typical examples of images for the largest fractal dimension  (and below and above) are shown in Figs. \ref{Figure:F_D_Dots} (b), (c) and (d)

\subsection{The Betti numbers of synthetic images}
\begin{figure}
\centering
\subfigure[]{\includegraphics[width=0.5\linewidth]{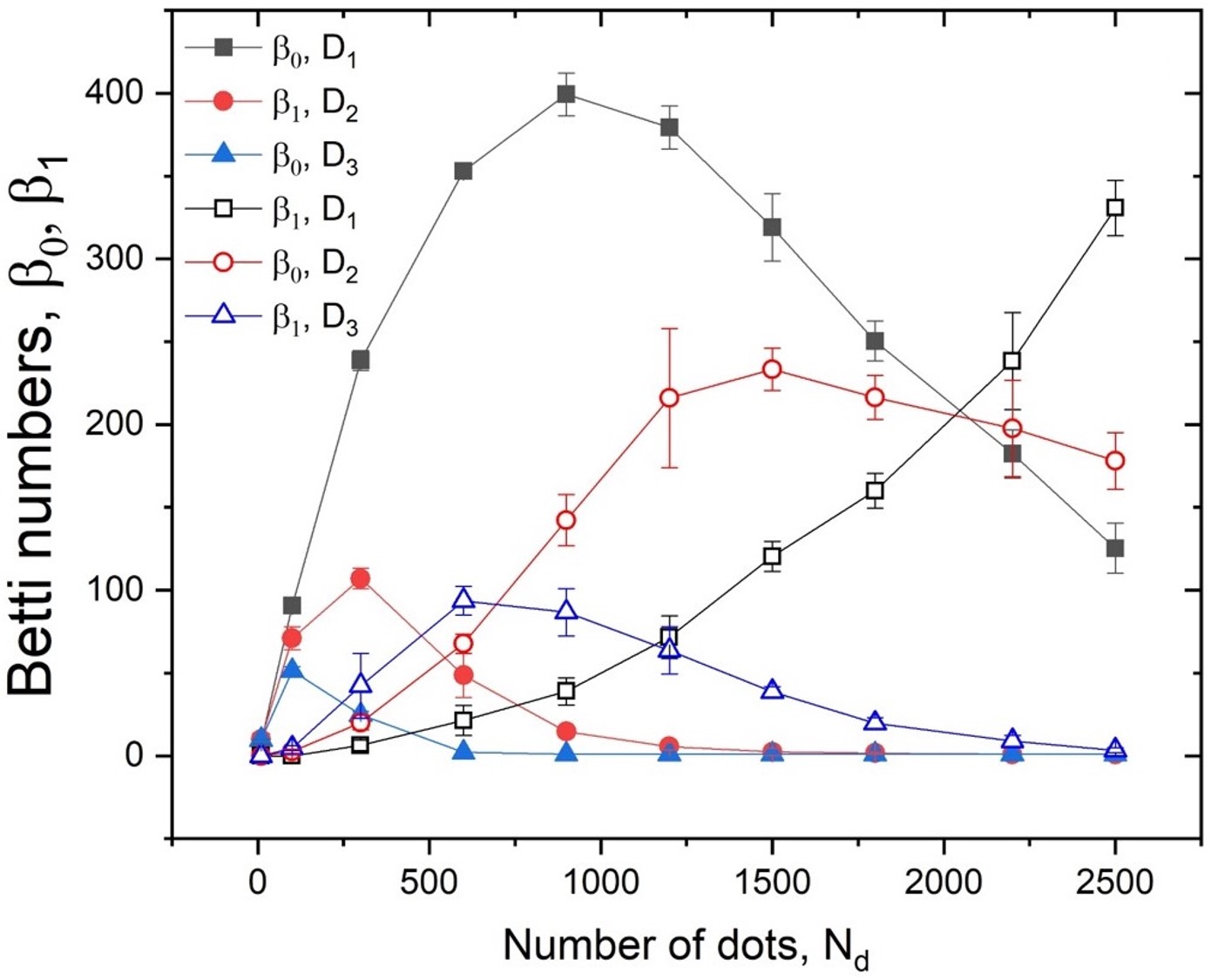}}\,
\subfigure[]{\includegraphics[width=0.5\linewidth]{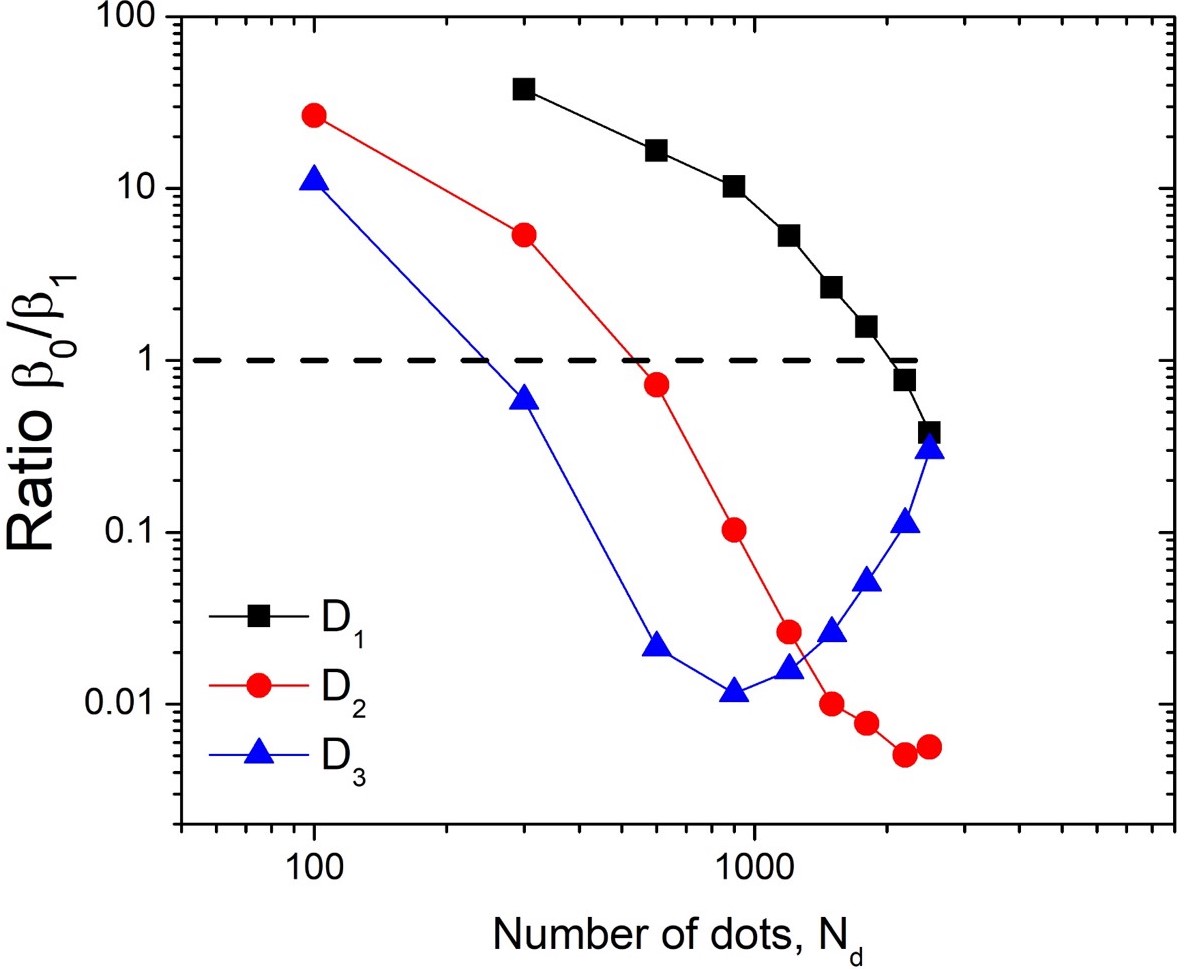}}\\
\subfigure[$\beta_0/\beta_1 > 1, N_d=200$]{\includegraphics[width=.3\linewidth]{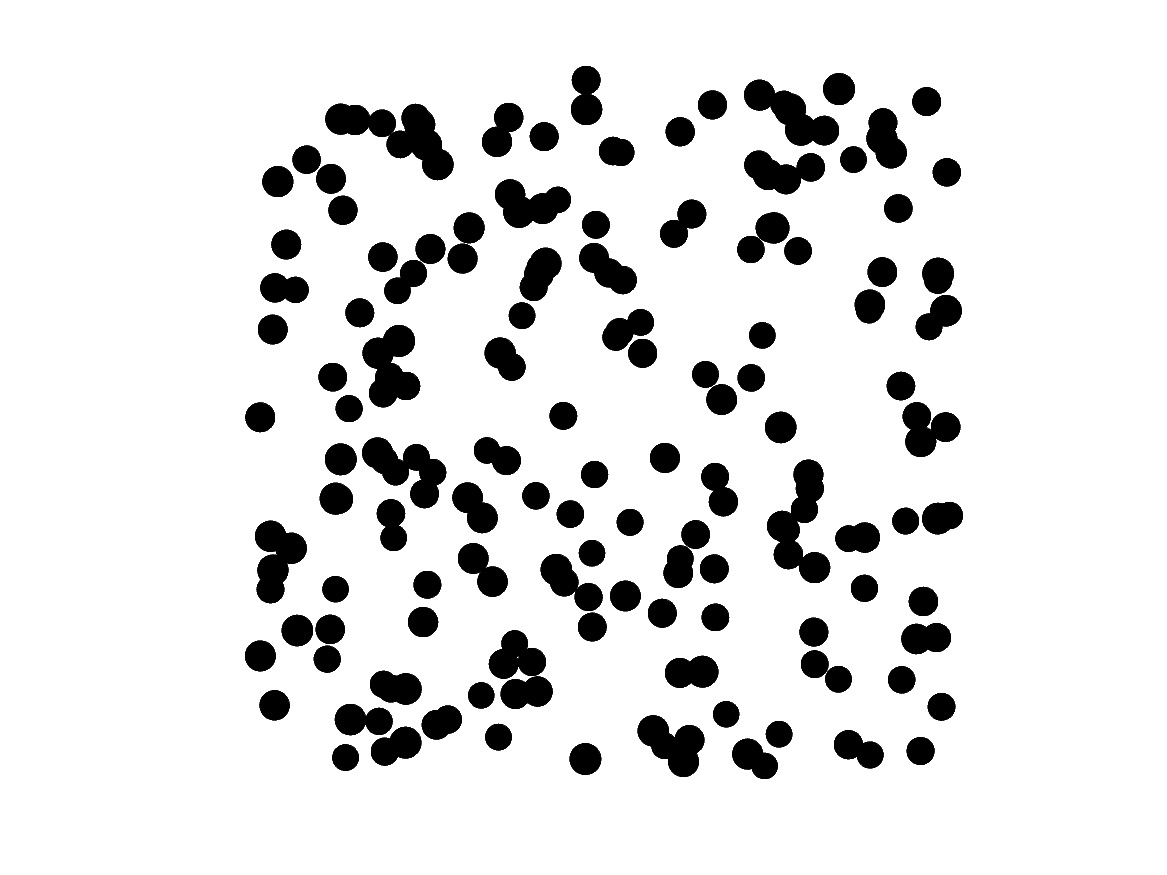}}\,
\subfigure[$\beta_0/\beta_1\approx1, N_d=500$]{\includegraphics[width=.3\linewidth]{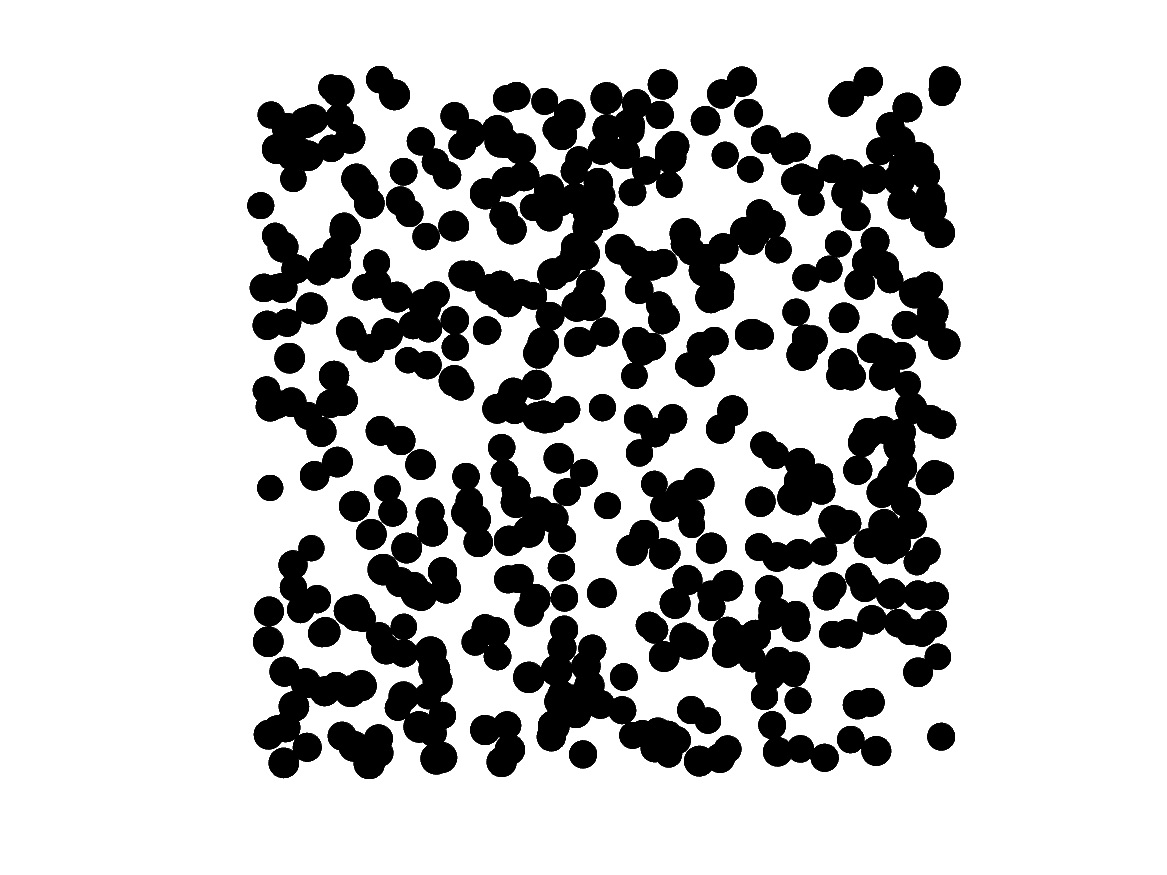}} \,
\subfigure[$\beta_0/\beta_1 < 1, N_d=1500$]{\includegraphics[width=.3\linewidth]{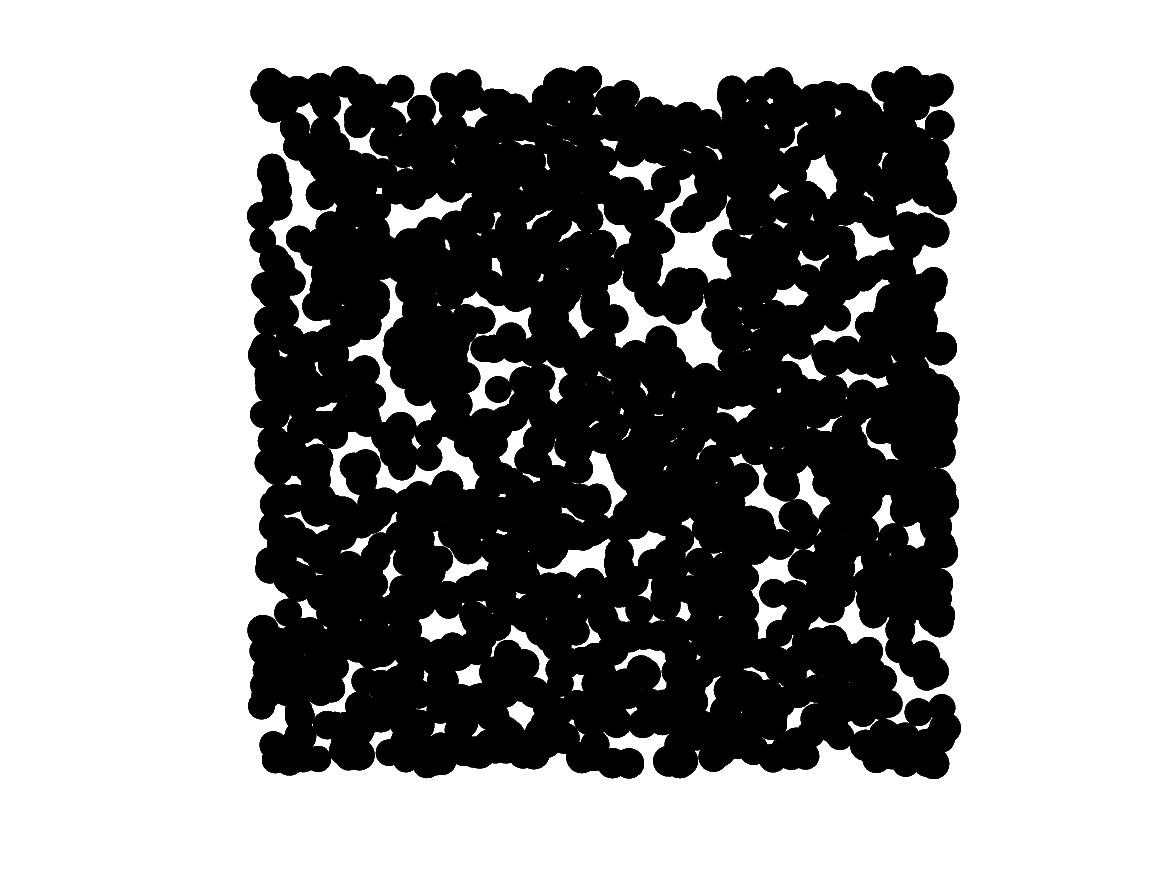}}
\caption{(a) The value of $\beta_0$ (fill symbols) and $\beta_1$ (empty symbols) as a function of $N_d$; (b)  ratio $\beta_0/\beta_1$  as a function of $N_d$ for the three different diameters $D_1$, $D_2$, $D_3$ (black squares, red circles and blue triangles respectively); (c,d,e) typical images showing different values of $\beta_0/\beta_1$ for $D_2$=0.02. }
\label{Figure:bettis_dots}
\end{figure}

The complexity characterized by the value of the Betti numbers is displayed in Fig. \ref{Figure:bettis_dots}(a). The value of the $\beta_0$, shown as filled symbols in the figure,  first increases with the number of dots since it measures the topological connectivity. As more dots are added, each new dot represents a simple-connected region, which indicates the linear increase of $\beta_0$. As $N_d$ increases overlap begins to occur, and the new dots do not create new simply-connected regions; hence, the rate if increased of $\beta_0$ slows down until a maximum value is observed. The maximum depends both on the number of drops and the the drop diameter. As the area of the image is progressively covered by drops, the value of $\beta_0$ decreases:  the new dots begin to fill space that connects regions leading to a decrease of $\beta_0$. When the dots fill almost completely the area of interest, the number of connections is again $1$ that is counted as one continuous set. For larger drops, the area is filled with fewer dots and the maximum is reached quickly. Conversely, the number of holes, quantified by the value if the $\beta_1$ number (empty symbols in the figure) starts from zero. When the dots are sparse and no overlaps occur, the simply-connected regions do not have any voids leading to a small value of $\beta_1$. But as overlaps become more frequent as the number of dots increases, some holes appear resulting in higher $\beta_1$ values.  As even more dots are added, the number of holes reaches a maximum value because too many overlaps will lead to the new dots filling up the holes. As with $\beta_0$ number, when the dots fill the area of interest, $\beta_1$ number again approaches $0$.

De la Calleja \emph{et. al} \cite{DelaCalleja:2017} argued that images for which $\beta_0\approx\beta_1$ had the maximum complexity. In other words, the images had as many connected regions as holes. To evaluate this criteria for the present images, Fig. \ref{Figure:bettis_dots} (b) shows the ratio $\beta_0/\beta_1$ as a function of $N_d$. In the figure, the ratio $\beta_0/\beta_1$ decreases as $N_d$ increases crossing a value of 1 at a certain number of dots, that depends on the size of dot diameter. As $N_d$ increases, the ratio contonues to decrease; for much large $N_d$ values the ratio appears to increase again.. Typical images for $\beta_0/\beta_1$ smaller, equal and larger than unity are shown in Figs.  Fig. \ref{Figure:bettis_dots} (c), (d) and (e), respectively. At first sight, the visual appearance of the images and their apparent complexity seems to be in agreement with the images shown in Figs. \ref{Figure:F_D_Dots} (b), (c) and (d), as the fractal dimension crosses through a maximum value. However, the number of dots at which the maximum fractal dimension occurs and at which $\beta_0/\beta_1\approx1$ is not the same. Also, the ratio $\beta_0/\beta_1$ evolves from a high to small value, crossing through the point of maximum complexity, while maximum complexity from the fractal dimension is located at the maxium of $f(\alpha)$.

A similar analysis was conducted for the case of straight lines, extending the study first used by De la Calleja \emph{et. al} \cite{DelaCalleja:2017}. The results for the case of lines hold the same qualitative behavior but the values at which the maximum complexity is reached depend on the length and thickness of the lines. These results are presented  in detail and discussed in  Appendix A.


\section{Complexity of Yayoi Kusama's infinity net}
Yayoi Kusama is an international avant-garde Japanese artist. Her `net painting' style has been deemed as her  most prototypical obsessional art~\cite{Pollock:2006jd}. Some critics have interpreted her style as `culturally specific', `historically feminine' and `subjective painting without image'~\cite{Pollock:2006jd}. The art of the so-called queen of \emph{Polka Dots} is characterized by constructions of by consecutive dots. The world-wide success of her recent \emph{Infinity Rooms}, has made Yayoi Kusama one of the most recognizable and admired contemporary artists.

Kusama's paintings have evolved significantly from simple stylistic observations in the 1950s to pure and personal expression of her mental illness in the 1990s \cite{Nakajima:2020dg}. Technically, Kusama's brushstrokes are very thick and  do not seem to finish.

Considering the technique describe above, we analyse Kusama's artworks in a more objective manner. 
In order to quantify the complexity of Kusama's \emph{infinity net} style, $30$ paintings were selected. A complete list of these works is included in Appendix B. Most of the artworks used here are from the following collections: \emph{Dot's obsession} \cite{Dot-obsession:2019dg}, \emph{Infinity Nets} \cite{Infinity-nets:2019dg}, \emph{My eternal soul} \cite{Kusama:2017qr} and \emph{Silkscreens from 2009} \cite{Silkscreens:2019dg}. The specific images were selected by considering the appearance of the characteristic repeated dot pattern and the absence, as much as possible, of figurative expressions or objects. The analysis of complexity in pieces of art is possible because of the two dimensional nature of the paintings. Although it is, in principle, possible to extend our proposal to 3-dimensional spaces it was not attempted. 

\begin{figure}
\begin{center}
     \includegraphics[width=0.8\textwidth]{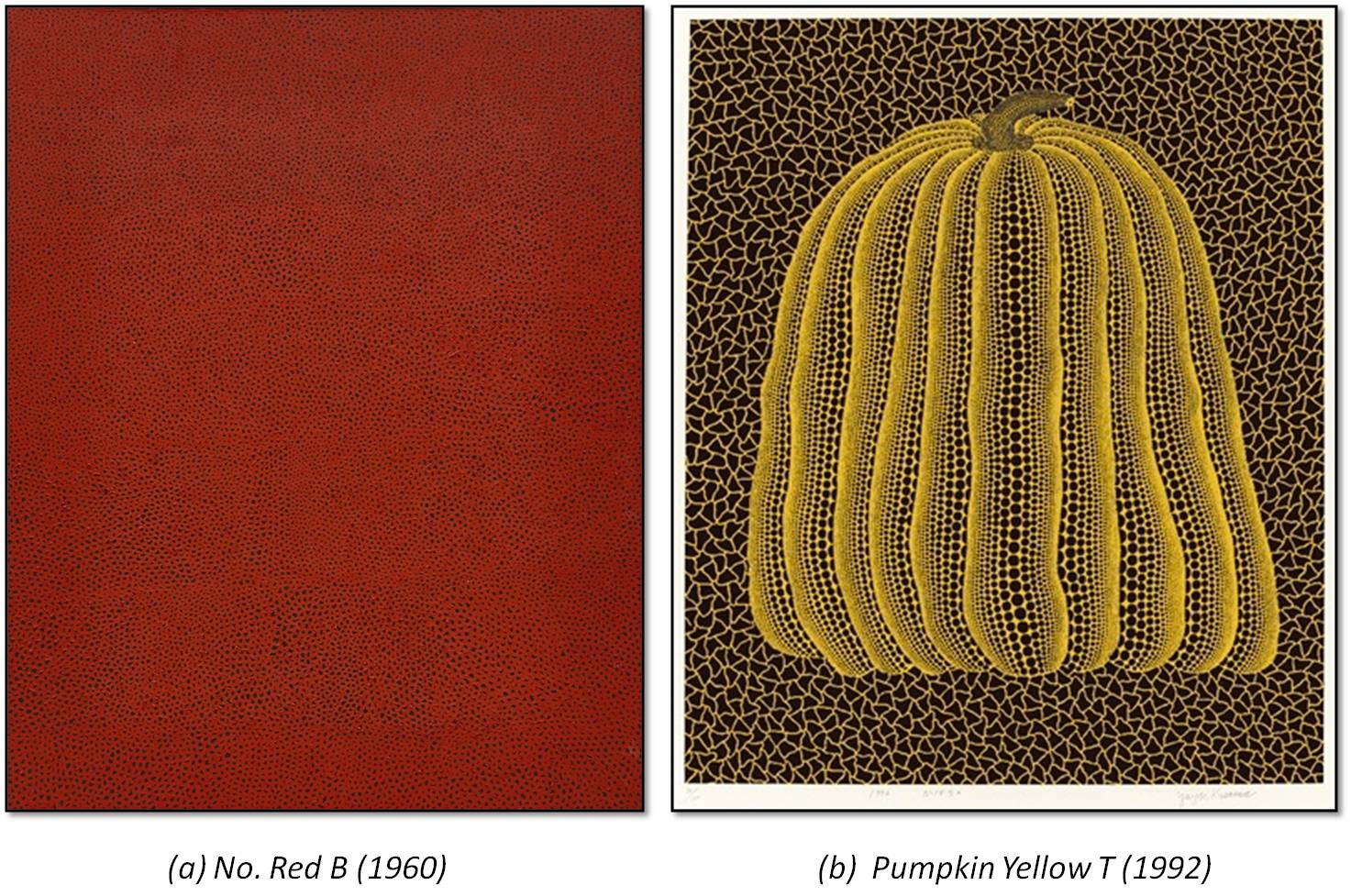}\\
     \includegraphics[width=0.88\textwidth]{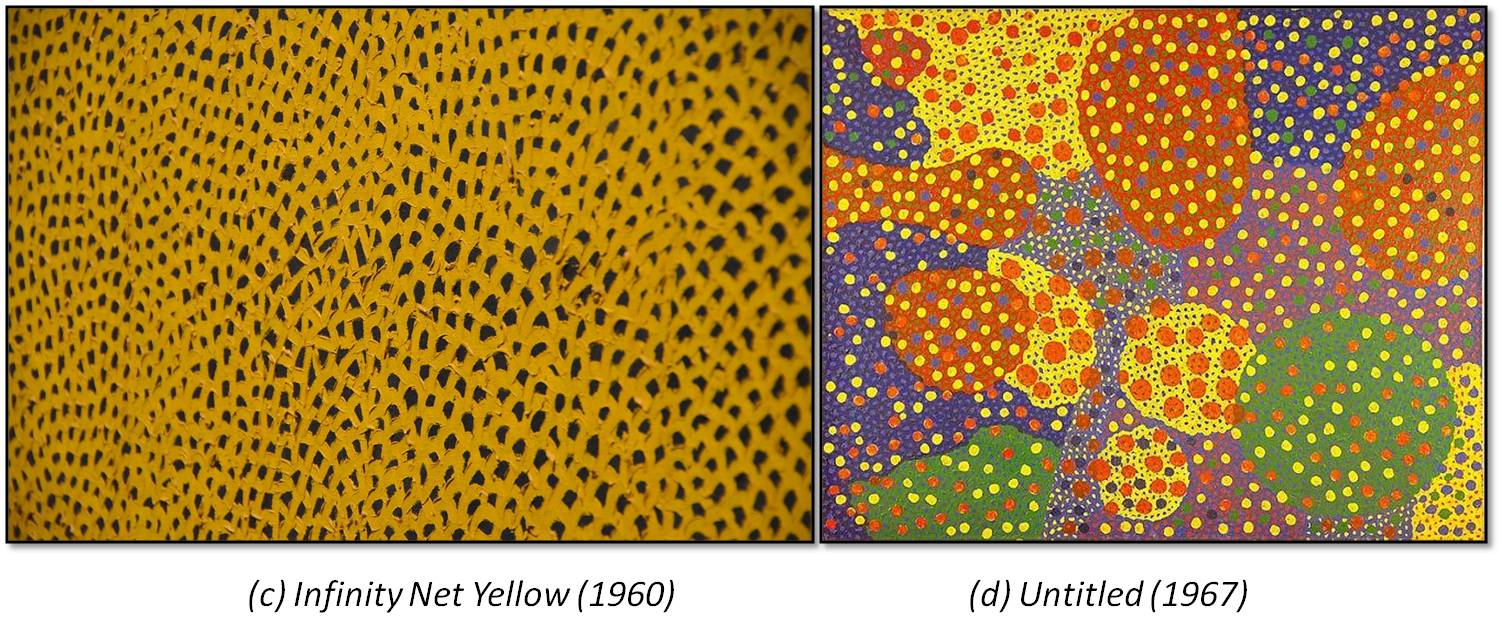}
\caption{Four of the most representative Kusama's paintings. $(a)$ No. Red B (1960), $(b)$ Pumpkin Yellow T (1992), $(c)$ Infinity Nets Yellow (1960) and, $(d)$ Untitled (1967). Images used with permission from the Yayoi Kusama Foundation.}
\label{Figure_Kusama-Artworks}
\end{center}
\end{figure}

Figure~\ref{Figure_Kusama-Artworks} show four of the most iconic Kusama's artworks: $(a)$ No. Red B (1960), $(b)$ Pumpkin Yellow T (1992), $(c)$ Infinity Nets Yellow (1960) and, $(d)$ Untitled (1967) \cite{kusama-REF:2019dg}.  The pieces of art of Kusama are full of the symmetric and irregular distribution of dots which apparently are a result of simple compositions. However, our results provide compelling evidence that this fact is questionable. 

\begin{figure}
\begin{center}
\includegraphics[width=0.7\textwidth]{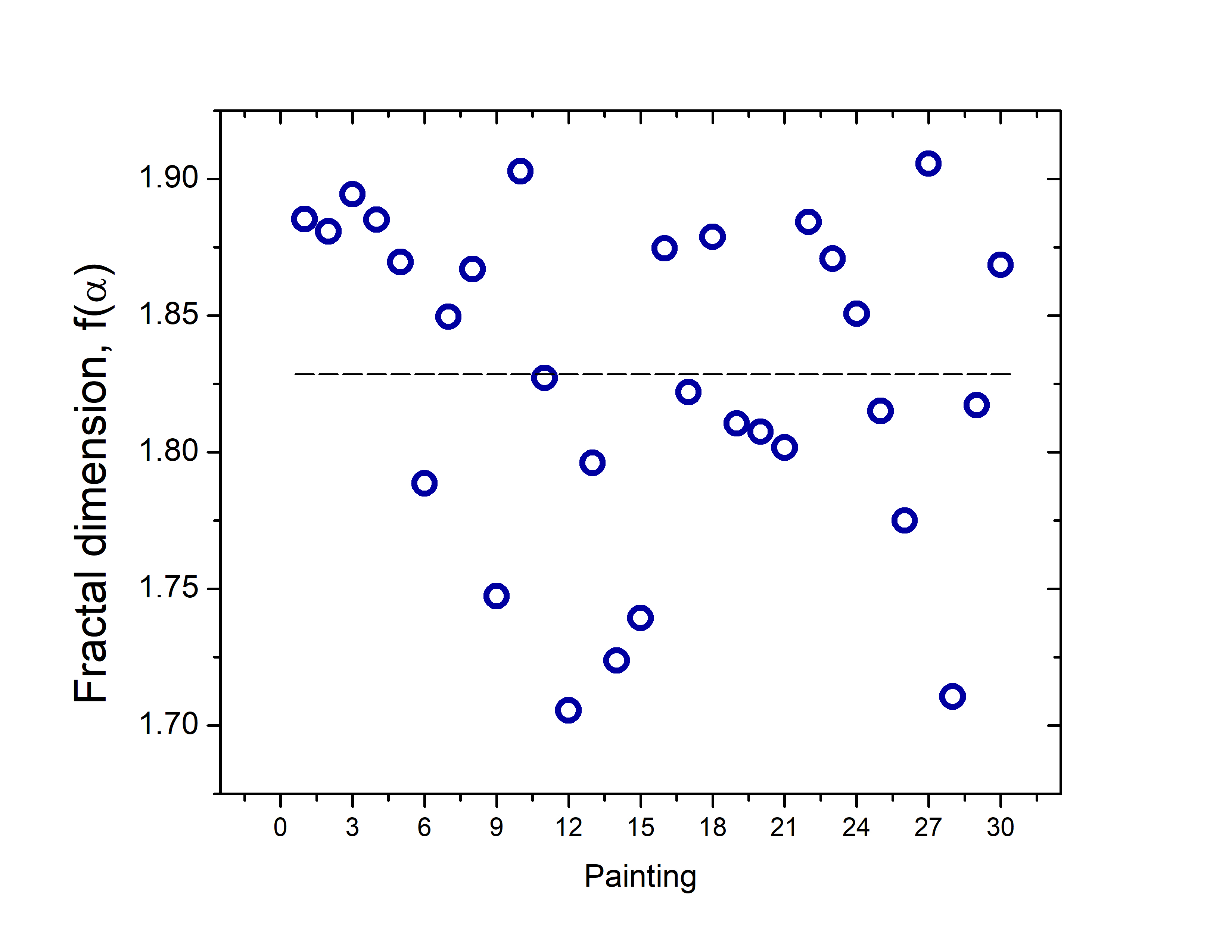}\\
\caption{The fractal dimension of Kusama's paintings. The paintings considered are shown in Tables \ref{Table:Kusama1} and \ref{Table:kusama2} in Appendix B.  The dashed line shows the mean value of $f(\alpha)$}
\label{Figure:DF_Kusama}
\end{center}
\end{figure}

The fractal dimension $f(\alpha)$ for the selected set of paintings is shown in Figure \ref{Figure:DF_Kusama}. As can be observed, the fractality of the Kusama's artworks is between $1.7<f(\alpha)<1.9$, with a mean value of $f(\alpha) = 1.8285$ (dashed black line). These range of fractal dimension is in good correspondence with other abstract images with a high density of dots with the smallest diameter as we observed in Figure \ref{Figure:F_D_Dots}. Also, Kusama's artworks fractal dimensions can be compared with those for other abstract expressionisms \cite{Elsama:2016dg}. Interenstingly. the fractality of the art pieces of Yayoi Kusama evaluated in this report have values similar to those calculated for Jackson Pollock's paintings \cite{Taylor:1999dg,Elsama:2016dg}. This result suggests that the despite the obvious differences in style and composition the complexity of Kusama's paintings,  using this metric,  is similar to that of Pollock's. 

Now, to complete the complexity characterization of Kusama's paintings we proceed to calculate the values of the Betti numbers. To do this, the images are binarized choosing a threshold level following ordinary image processing techniques\cite{Schneider:2012dg}. The value of $\beta_0$ number quantifies the connectivity of a set and, while the value of $\beta_1$ number counts the holes. Figure \ref{Figure:Kusama_Bettis} (a) show the Betti numbers for Kusama's paintings listed in Appendix B. Note that both $\beta_0$ and $\beta_1$ are normalized with the area of each painting (A), to allow comparison between works of different sizes. The mean value of $\beta_0$  (brown dashed line) and the mean value of $\beta_1$ number (green dashed line) are also shown. In general, we find that $\beta_0>\beta_1$
There is expected since the Kusama's dots are mostly disconnected, each one of them represents a single connected domain with would lead to a high $\beta_0$ value. Since there are few overlaps, only a few holes appear, therefore one would expect the images to have low  $\beta_1$ values. This, of course, is not always the case. In many instances the  high density of dots are in balance with the number of holes.

We have conjectured that a high image complexity corresponds to cases when the number of connected regions is roughly the same as the number of holes, $\beta_0 \sim \beta_1$ \cite{Elsama:2016dg}.  Fig.\ref{Figure:Kusama_Bettis}$(b)$ shows the ratio $\beta_0/\beta_1$ of all the works analyzed here. Overall, it is clearly observed that $\beta_0 > \beta_1$. This is good agreement with can be deduced from the visual inspection of  the paitings: dispersed non-overlapping points do not lead to the formation of many holes. Of all the works, only 1/3 of them a value of $\beta_0/\beta_1$ close to unity. Hence, from this metric the complexity of Kusama's paintings seems to be smaller. This observation is in sharp contrast with what was concluded from the fractal dimension analysis. It is also interesting to note that for the case of Pollock's paintings, the ratio $\beta_0/\beta_1$ was closed to unity, at least for works close to the dripping period. 
\begin{figure}
\begin{center}
\centering
\subfigure[]{\includegraphics[width=0.5\textwidth]{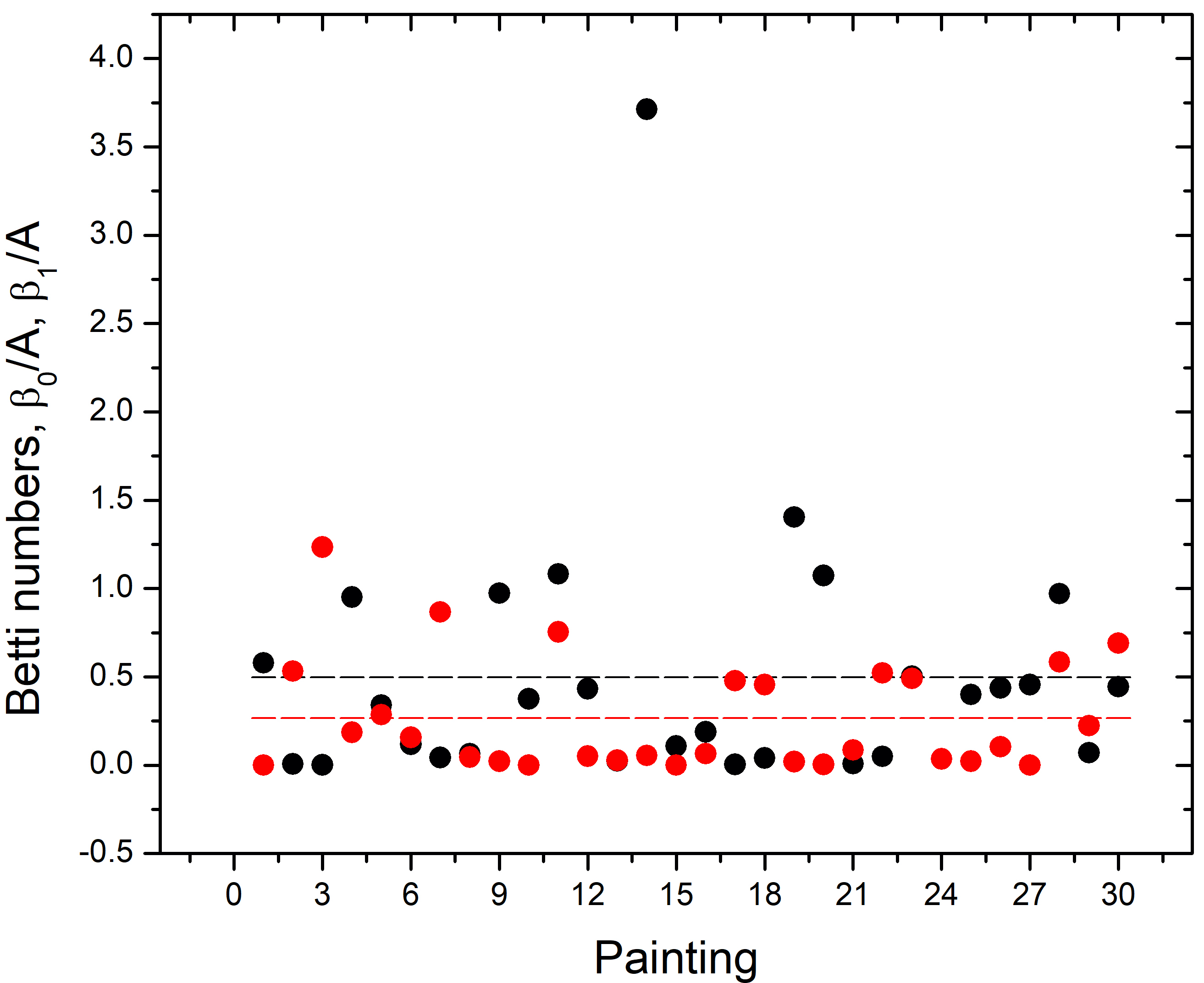}}\\
\subfigure[]{\includegraphics[width=0.5\textwidth]{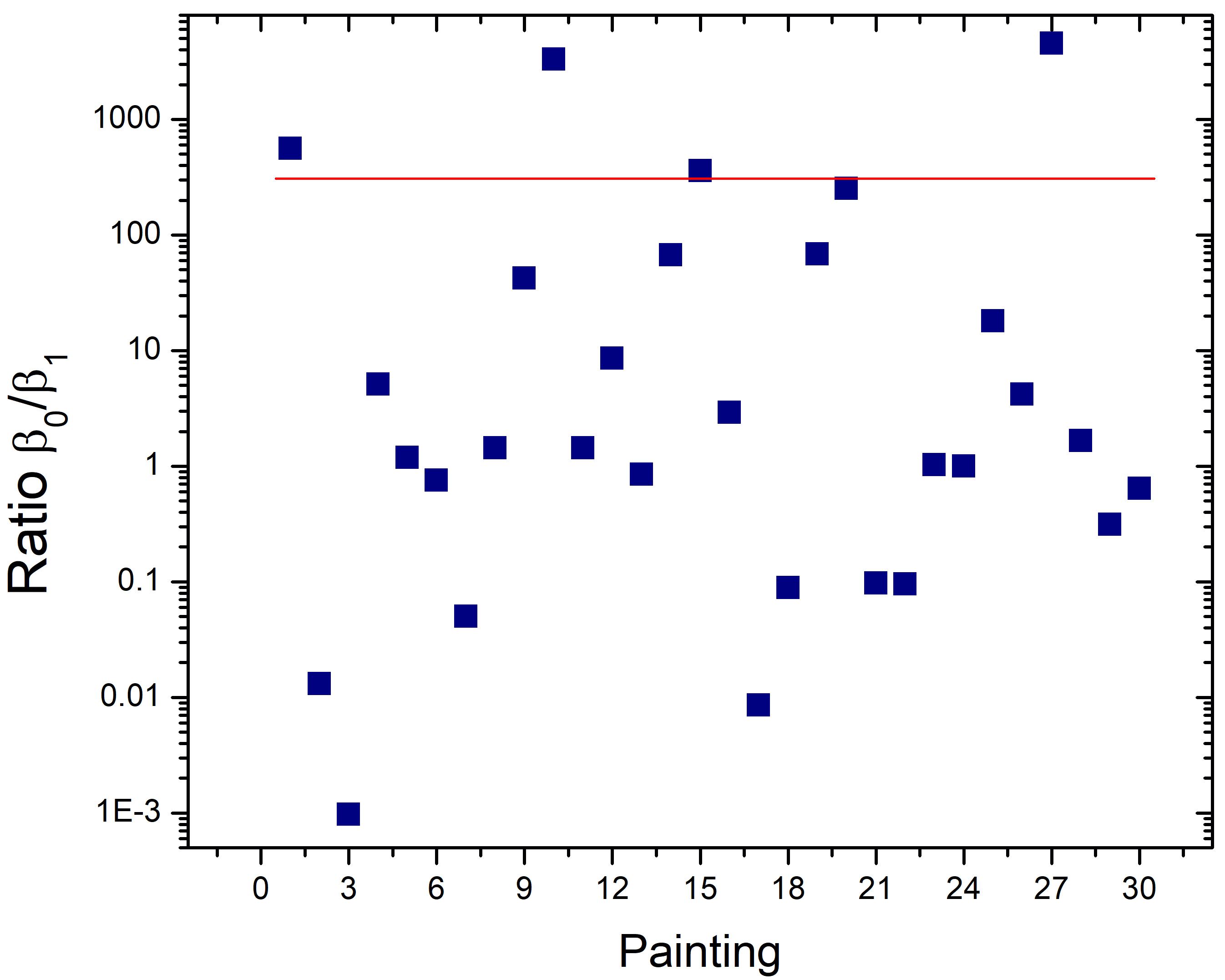}}\\
\caption{(a) Betti numbers per unit area of Kusama's paintings, list in Appendix B. Black circles: $\beta_0$; red circles, $\beta_1$. The black dashed line shows the mean value of $\beta_0/A$ and the red dashed line show the mean value of $\beta_1/A$. $(b)$ the ratio $\beta_0/\beta_1$ for the same paintings. The solid line shows the mean value.}
\label{Figure:Kusama_Bettis}
\end{center}
\end{figure}


\section{Discussion and Conclusion}
The fractal dimension and Betti numbers can both be used  to characterize complexity with a non-predictable structure. Fractal properties are present in a wide and extend the number of objects. While the fractal dimension measures certain spectral properties of an image, the Betti numbers quantify the topological shape. As demonstrated above, they can both provide a metric for complexity of abstract images. One interesting aspect conclusion drawn from this study was that, although Kusama's art works have fractal dimensions comparable to other abstract paintings, the value of the Betti numbers are strikingly different. 


The analysis of image complexity is a useful non-invasive technique to characterize artworks.  With the analysis presented here we demonstrate that it is possible to capture  essential information about painting styles. Yayoi Kusama has painted many pieces with dots, and her influence and impact have been significant.  Kusama's paintings are created with consecutive and order distribution of dots. From our analysis, the fractality of Kusama's art indicates high complexity while the value of the topological invariants show more homogeneous textures with a high degree of dispersiveness. Hence, we can conclude the complexity of Kusama's art work has a richer depth than that observed for other abstract painters, including Jackson Pollock.

\subsection{Acknowledgments}
The support of DGAPA-PAPIIT-UNAM (grant number IN108016) and ACT-FONCA (grant number 04S.04.IN.ACT.038.18) are greatly acknowledged. We thank the Yayoi Kusama foundation for their kind permission to use the images of the art works.




\bibliographystyle{unsrtnat}
\bibliography{topology}

\appendix

\section{Synthetic images with lines}
De la Calleja \emph{et. al} \cite{DelaCalleja:2017} reported the complexity of abstract images composed of rigid lines varying the length and the thickness. In the present  case, we are interested in the effect of changing the line thickness.  To evaluate the complexity  abstract images were generated using the same script written in Matlab. The procedure to generate images with rigid lines in the three different thickness was the same as the one used with dots. Black rigid lines were generated to fill a fixed black space. Three different thickness were computed $T_1=0.1$, $T_2=0.3$ and $T_3=0.5$. The lines were randomly distributed on the space. We follow the same methodology to evaluate the complexity of images generated with dots, but applied to images with lines. The fractal dimension ($f(\alpha)$) and Betti numbers were calculated.

\begin{figure}[!ht]
\centering
\includegraphics[width=0.85\linewidth]{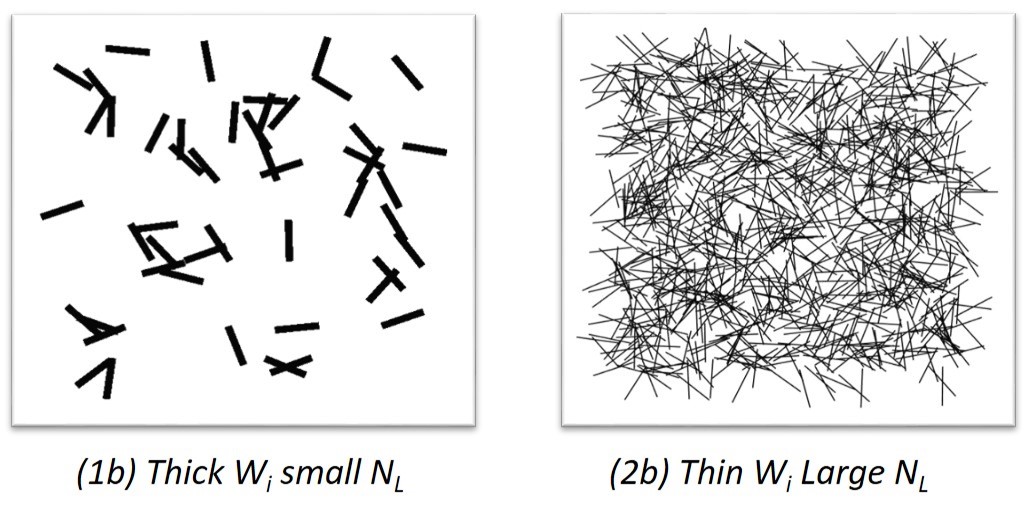}
\caption{Typical synthetic abstract images generated with random distribution of lines.}
\label{Image:abstract_images_examples}
\end{figure}

\begin{figure}
\begin{center}
\includegraphics[width=0.78\linewidth]{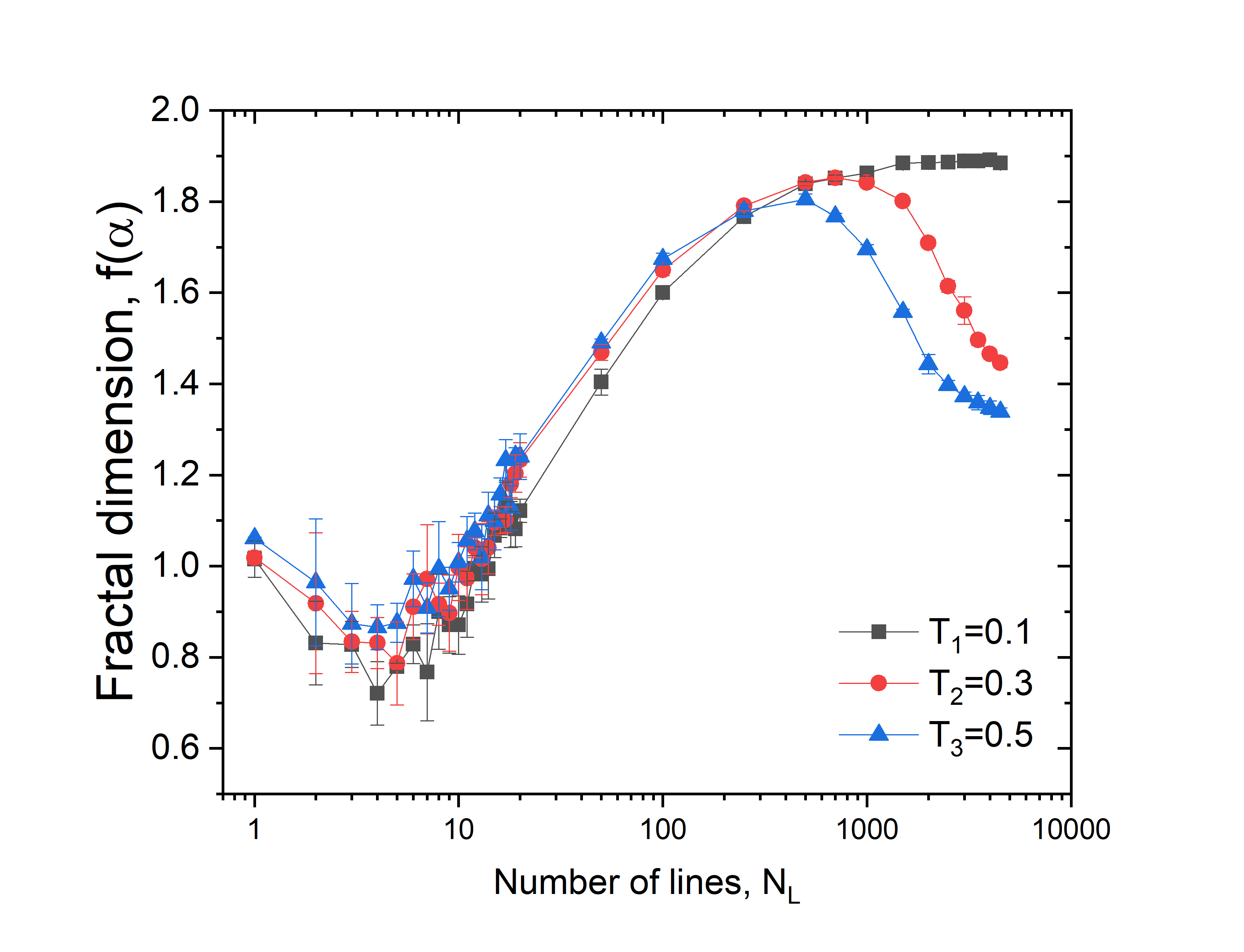}
\caption{The Fractal dimension $f(\alpha)$ reaches rapidly almost a value of $2.0$. It happens for the three thickness $T_1,T_2,T_3$ (black squares, red dots, and blue triangles respectively).}
\label{Figure:F_D_Lines}
\end{center}
\end{figure}

\begin{figure}
\begin{center}
\includegraphics[width=0.8\linewidth]{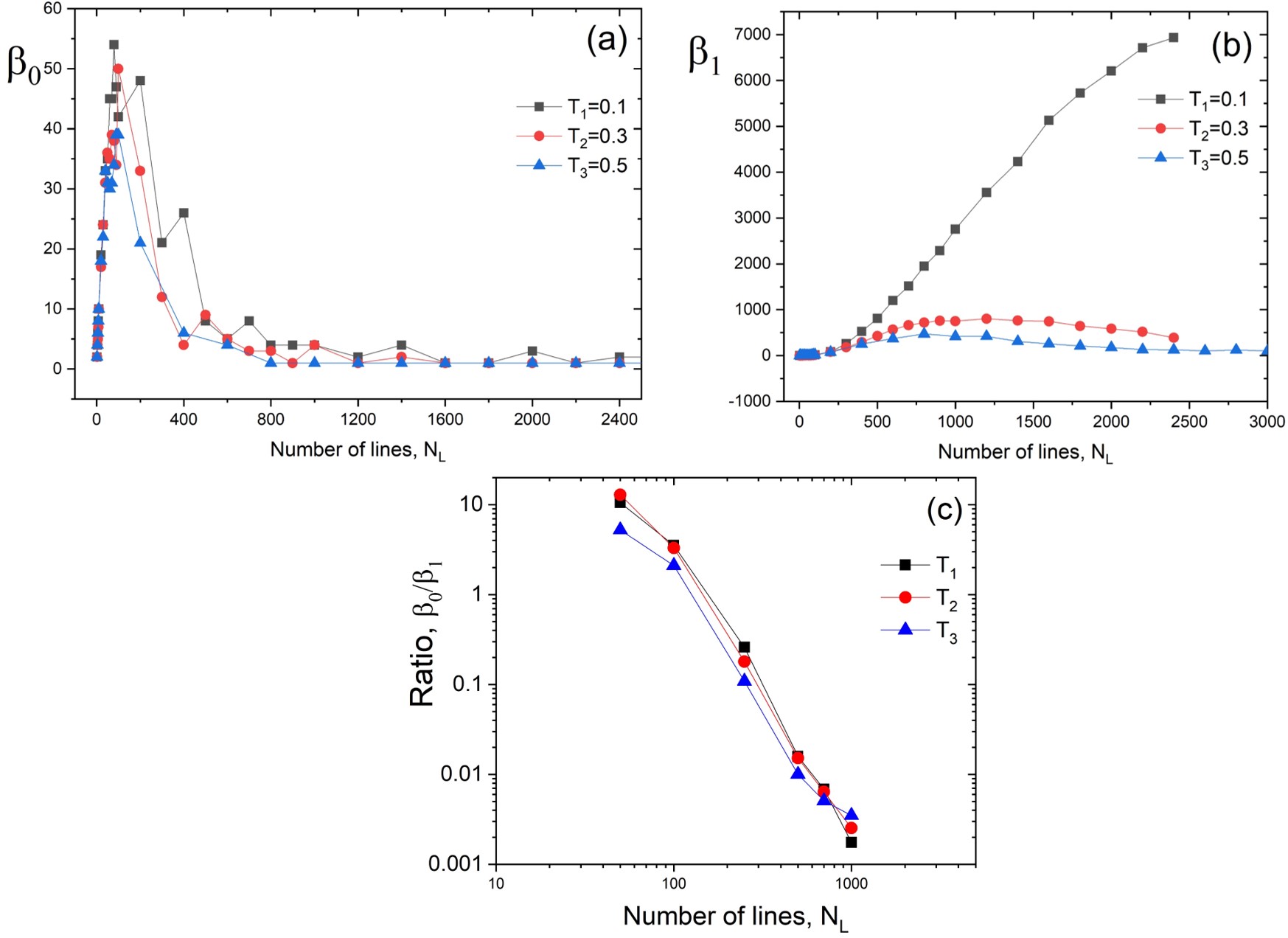}\\
\caption{In $(a)$ is presented that high connectivity measured with $\beta_0$ on abstract images generated with lines is reached with a low number of lines on the three different thickness: $T_1$, $T_2$ and $T_3$ (black squares, red dots, and blue triangles respectively). $(b)$ exhibits the complexity measured by the number of holes obtained from the overlapped lines on the images. Images constructed with thin lines (black squares) exhibit the highest value of $\beta_1$. In (c) the ratio $\beta_0/\beta_1$ of abstract images generated with lines of three different thickness shows a decrease of complexity when the images contain a high number of lines.}
\label{Figure:Bettis_Lines}
\end{center}
\end{figure}

The fractal dimension $f(\alpha)$ as a function of the number of lines ($N_L$) is presented in Fig. \ref{Figure:F_D_Lines}.  The fractal dimension start from a value of unity for a single line, as expected. As more lines are added, the dimension decreases slightly to soon after increase again. As more lines are added, the space is filled rapidly which leading to a  maximum fractal dimension of about 1.8. This is observed for the the three thicknesses, but the maxium is reached at different $N_L$. After the maximum value of fractal dimension is reached, which means that the empty space is fill with lines the fractal dimension decreases agian. 
The Betti numbers obtained for the same images are shown in Fig. \ref{Figure:Bettis_Lines}. The results of $\beta_0$ of images with lines are presented in $(a)$. The connectivity of lines is reached rapidly with a low number of lines and this behavior is observed on images computed with the three thicknesses. $\beta_0$  collapse to almost $1$ from images with $2000$ lines. This is because as the number of lines increases, they connected and overlapped between them and form one set. In  Fig. \ref{Figure:Bettis_Lines} (b)  the behavior of $\beta_1$ is shown, which measures the number of empty spaces of a set. Images generated with thin lines ($T_1$) reached a high number of holes. Contrary to the cases when the lines are thick. In those cases, the empty space is filled rapidly with a low number of lines. We have argued that a high level of complexity is obtained when $\beta_0 \sim \beta_1$. In Fig. \ref{Figure:Bettis_Lines} (c) the ratio $\beta_0/\beta_1$ is shown as a fucntion of $N_L$. It shows that the maximum complexity obtained from images with three different thicknesses occurs at approximately the same number of lines $N_L\approx 100$ independently of the line thickness. 

\newpage

\section{Kusama's paintings}
The body of work by Yayoi Kusama is vast. Many of her paintings have been exposed in important galleries around the world, including performances and pieces  were the public can interact with the artwork. Among them `Net painting' has been considered to be her most prototypical example of `obsessional art'. 

In Yayoi Kusama's own words:  `My life, a dot, namely, one among millions of particles. It was in 1959 that I gave my manifesto that [my art] obliterates myself and others with the void for a net woven with an astronomical accumulation of dots'. Since the end of the 1950s, Kusama has produced a many artworks under different titles, different sizes, color schemes but with an apparently repetitive pattern: infinity nets.

The `self-obliteration' of Kusama's words appear as the extreme opposite of the `action painting'. During the painting process, her brush does not leave the canvas, in sharp contrast to Pollock's free-hand  action \cite{Zenit:2019dg}.

At first sight, Kusama's paintings may appear to be simple monotonous patterns of a net. However, the monotonous look may prompt various associations with the obsessions of the artist. The ideological, social, and physiological aspects of Kusama's art, have been discussed in diverse forums. Kaniichi \cite{Uno:1988dg} discusses the negation of action painting and posses the question: can the practice of `Net' production be a feminine alternative to the masculine staging of `action' as she appeared to produce the `opposite' of Pollock type paint? 

In \emph{Net painting}, the act of painting is austerely restrained to minimal action. If action painting is a trace of a progressive `cutting off' of the body of the painting, net painting might suggest a `touching' without harming. Pollock's action is oriented to detach the artist form the body of the painting; Kusama's continuous painting aspires to produce contiguity with her object \cite{Nakajima:2020dg}. 

our results show that despite having contrastingly different styles and approach to painting, Kusama and Pollock have similar values of complexity evaluated with fractal dimension. However, if are the Betti number the parameter of characterization, Kusama's work shows low level of complexity. 

\begin{table}[ht]
\scalebox{0.7}{
\begin{tabular}{|c|c|c|c|l|}
\hline
Paint & Year & Name  & Size  & \multicolumn{1}{c|}{Source} \\ \hline
1     & 1960 & White No. 28 & 147.6x111.1 & \begin{tabular}[c]{@{}l@{}}https://www.christies.com/\\lotfinder/Lot/yayoi-kusama-b-1929- \\ white-no-28-5846063-details.aspx\end{tabular} \\ \hline
2     & 1998 & Sunlight  & 59.7x47.8   & \begin{tabular}[c]{@{}l@{}}http://www.artnet.com/artists/\\ yayoi-kusama/sunlight-EzrsqLuca\\ Tdm45F84jNNgA2\end{tabular}  \\ \hline
3     & 1960 & Infinity nets & 73x91.1   & \begin{tabular}[c]{@{}l@{}}https://www.artsy.net/artwork/\\yayoi-kusama-infinity-nets-1960-1  \end{tabular} \\ \hline
4     & 1992 & Pumpkin (Yellow T)  & 72.3x60.4   & \begin{tabular}[c]{@{}l@{}}https://www.artsy.net/artwork/\\ yayoi-kusama-pumpkin-yellow-t\end{tabular}   \\ \hline
5     & 1988 & Petals  & 53x45  & \begin{tabular}[c]{@{}l@{}}https://www.phillips.com/detail/\\ yayoi-kusama/UK010612/143\end{tabular}  \\ \hline
6     & 1988 & \begin{tabular}[c]{@{}c@{}}Waves on the\\ Hudson River\end{tabular}                        & 45.5x38     & \begin{tabular}[c]{@{}l@{}}https://www.mutualart.com/Artwork/\\ Waves-on-the-Hudson-River/\\ 3D6A54B3E798DCD0\end{tabular}  \\ \hline
7     & 2009 & \begin{tabular}[c]{@{}c@{}}Night Ripples \\ {[}TOWSS{]}\end{tabular}  & 130.3x162   & \begin{tabular}[c]{@{}l@{}}https://www.roslynoxley9.com.au/\\ artwork/yayoi-kusama-\\ night-ripples-towss-0/33-14041\end{tabular}  \\ \hline
8     & 2006 & A Flowing River  & 162x130.3   & \begin{tabular}[c]{@{}l@{}}https://www.roslynoxley9.com.au/\\exhibition/selected-works-\\silkscreens/vyft4\end{tabular}  \\ \hline
9    & 1953 & Waves  & 26.4x33     & \begin{tabular}[c]{@{}l@{}}http://www.artnet.com/artists/\\ yayoi-kusama/waves-\\ W2Lf-D1xCrZ\_Wy1IEBlFrg2\end{tabular}    \\ \hline
10    & 1960 & \begin{tabular}[c]{@{}c@{}}Yellow net \\ (Infinity) \end{tabular} &    240x294.6         & \begin{tabular}[c]{@{}l@{}} \\ https://www.nga.gov/collection \\ /art-object-page.124183.html\end{tabular}    \\ \hline
11    & 2011 & Infinity nets LNXA  & 97x130.3    & \begin{tabular}[c]{@{}l@{}}https://www.phillips.com/detail/\\ YAYOI-KUSAMA/UK010612/139\end{tabular}    \\ \hline
12    & 1986 & Mountain country  & 38x45.7     & \begin{tabular}[c]{@{}l@{}}http://www.barbaramathesgallery.\\ com/exhibition/yayoi-kusama-\\ from-here-to-infinity/selected-works/3\end{tabular}   \\ \hline
13    & 1992 & \begin{tabular}[c]{@{}c@{}}The sky in the\\  evening glow \\ (Red rain) \end{tabular}       & 161.9x227.3 & \begin{tabular}[c]{@{}l@{}} https://guyhepner.com/product/\\rain-evening-glow-yayoi-kusama/ \end{tabular}  \\ \hline
14    & 1988 & Passage of the wind     & 53.3x45.7   & \begin{tabular}[c]{@{}l@{}}https://www.artsy.net/artwork/\\ yayoi-kusama-passage-of-the-wind\end{tabular}  \\ \hline
15    & 2011 & Red dots   & 100x100     & \begin{tabular}[c]{@{}l@{}}https://www.christies.com/\\ lotfinder/Lot/yayoi-kusama-b-1929-\\red-dots-6191623-details.aspx\end{tabular}  \\ \hline
\end{tabular}
}
\caption{List of Yayoi Kusama's art works used in this investigation, part I.}
\label{Table:Kusama1}
\end{table}

\begin{table}[ht]
\scalebox{0.7}{
\begin{tabular}{|c|c|c|c|l|}
\hline
Paint & Year & Name  & Size  & \multicolumn{1}{c|}{Source} \\ \hline
16    & 2017 & \begin{tabular}[c]{@{}c@{}}I want to live\\  forever\end{tabular}                          & 194x194     & \begin{tabular}[c]{@{}l@{}}https://www.cobosocial.com/dossiers/\\yayoi-kusama-a-dot-in-the-universe/\end{tabular}  \\ \hline
17    & 2009 & \begin{tabular}[c]{@{}c@{}}Standing on the\\  riverbank of my\\  hometown\end{tabular}     & 145.5x112   & \begin{tabular}[c]{@{}l@{}}https://www.victoria-miro.com/\\ exhibitions/427/works/f05e6aa9496d58\end{tabular}   \\ \hline
18    & 2007 & \begin{tabular}[c]{@{}c@{}}Birth, Ageing, \\ Sickness and Death\\ {[}QXPAT{]}\end{tabular} & 130.3x162   & \begin{tabular}[c]{@{}l@{}}https://ocula.com/art-galleries/ \\ victoria-miro-gallery/artworks/ \\ yayoi-kusama/birth-ageing- \\ sickness-and-death-qxpat//\end{tabular}  \\ \hline
19    & 2014 & \begin{tabular}[c]{@{}c@{}}Infinity dots \\ (EFY)\end{tabular}                             & 130.3x97    & \begin{tabular}[c]{@{}l@{}}https://ocula.com/art-galleries/\\ victoria-miro-gallery/artworks/ \\ yayoi-kusama/infinity-dots-(efy)\end{tabular} \\ \hline
20    & 2014 & \begin{tabular}[c]{@{}c@{}}Infinity-Dots\\ {[}HOFS{]}\end{tabular}                         & 130.3x97    & \begin{tabular}[c]{@{}l@{}}https://ocula.com/art-galleries/\\ victoria-miro-gallery/artworks/ \\ yayoi-kusama/infinity-dots-hofs/\end{tabular}  \\ \hline
21    & 1994 & Yellow threes   & 162.1x390   & \begin{tabular}[c]{@{}l@{}}https://www.fronterad.com/\\ yayoi-kusama-una-delicada-locura/\end{tabular}  \\ \hline
22    & 2008 & \begin{tabular}[c]{@{}c@{}}Infinity nets\\ {[}OWTWQB{]}\end{tabular}                       & 162x130.3   & \begin{tabular}[c]{@{}l@{}}https://ocula.com/art-galleries/\\ roslyn-oxley9/artworks/yayoi-kusama/\\infinity-nets- owtwqb/\end{tabular}  \\ \hline
23    & 1991 & The Galaxy  & 61x91.4     & \begin{tabular}[c]{@{}l@{}}https://www.phillips.com/detail/\\ yayoi-kusama/NY010117/20\end{tabular}  \\ \hline
24    & 2009 & \begin{tabular}[c]{@{}c@{}}Late night Chat\\ is Filled with \\ Dreams\end{tabular}         & 162x162     & \begin{tabular}[c]{@{}l@{}}http://amorartinfo.blogspot.com/2012/08/\\ yayoi-kusama-un-mundo-de-pesadillas.html\end{tabular}  \\ \hline
25    & 2001 & \begin{tabular}[c]{@{}c@{}}When the soul\\ bursts into flames\end{tabular}                 & 82x62.5x21  & \begin{tabular}[c]{@{}l@{}}https://www.roslynoxley9.com.au/\\ artists/49/ $Yayoi_Kusama$/516/39817/\end{tabular}  \\ \hline
26    & 1967 &  Untitled & 101.6x127  & \begin{tabular}[c]{@{}l@{}}https://www.artsy.net/article/\\artsy-editorial-6-works-explain-yayoi-\\kusamas-rise-art-stardom\end{tabular}  \\ \hline
27    & 2013 &  Infinity Nets PEAA & 100X100  & \begin{tabular}[c]{@{}l@{}}https://www.artsy.net/artwork/\\yayoi-kusama-infinity-nets-peaa\end{tabular}  \\ \hline
28    &  1967 & Untitled  & 40.6x45.7  & \begin{tabular}[c]{@{}l@{}} https://ackland.org/files/2020/04/\\Kusama-Exhibition-Text-and-Images.pdf \end{tabular}  \\ \hline
29    & 2014 & All the eternal love  & 194x194  & \begin{tabular}[c]{@{}l@{}} https://www.briefltd.com/yayoi-kusama/ \end{tabular}  \\ \hline
30    & 2010 & Endless life of people   &  40.6x40.6  & \begin{tabular}[c]{@{}l@{}} https://www.artsy.net/artwork/ \\ yayoi-kusama-endless-life-of-people-6 \end{tabular}  \\ \hline
\end{tabular}
}
\caption{List of Yayoi Kusama's art works used in this investigation, part II.}
\label{Table:kusama2}
\end{table}

\end{document}